\documentclass[twoside]{IEEEtran}

\usepackage[dvips]{graphicx}
\usepackage{subfigure}
\usepackage{amsmath}
\usepackage{amssymb}

\newcommand{\tcm}{trellis-coded modulation}
\newcommand{\TCQ}{Trellis-Coded Quantization}
\newcommand{\tcq}{trellis-coded quantization}
\newcommand{\TCQr}{Trellis-Coded Quantizer}
\newcommand{\tcqr}{trellis-coded quantizer}
\newcommand{\hd}{Hamming-distance}
\newcommand{\ed}{Euclidean-distance}

\newcommand{\td}{two-di\-men\-sional}

\begin{document}
\title{{\TCQ} Based on Maximum-Hamming-Distance Binary Codes}

\author{Lorenzo~Cappellari,~\IEEEmembership{Member,~IEEE}%
\thanks{L.~Cappellari is with the Dept.~of Information Engineering of the University of Padova, Italy.}}

\maketitle

\begin{abstract}
Most design approaches for trellis-coded quantization take advantage of the duality of trellis-coded quantization with trellis-coded modulation, and use the same empirically-found convolutional codes to label the trellis branches. This letter presents an alternative approach that instead takes advantage of maximum-Hamming-distance convolutional codes. The proposed source codes are shown to be competitive with the best in the literature for the same computational complexity.

\end{abstract}


\IEEEpeerreviewmaketitle

\section{Introduction}
\IEEEPARstart{I}{t is} well known that trellis source coding asymptotically achieves the bounds of Shannon theory at the price of constructing (and storing) huge codebooks of random reproduction values \cite{viterbi_TrellisCoding}. More feasible trellis codes are instead obtained by reducing and structuring the codebooks. Trellis-coded quantization (TCQ) represents a successful attempt in this direction \cite{marcellin_TCQ}.

TCQ has first appeared as a natural counterpart, in the source coding framework, of {\tcm} \cite{ungerboeck_ChCodWithMultilevelSignals}. The basic TCQ scheme uses an extended codebook of $2^{R+1}$ reproduction values to code a source at a rate of $R$ bit/sample. This codebook is divided into four sub-codebooks that are assigned to the branches of a trellis diagram by means of a $(2,1)$ binary convolutional code, and the Viterbi algorithm is used to find the minimum mean-squared-error path through the trellis. Among the others, TCQ and TCQ-like coding (\cite{fisher_TCVQ,wang_TCVQ}) have found application in image coding \cite{taubman_JPEG2000,bilgin_ProgressiveTCQ}, in scalable coding \cite{wang_succRef-tcq} and in multiple description coding \cite{wang_md-tcq}.

Since the choice of the actual reproduction values is responsible for the achievable performance \cite{pearlman_codingUsingQuantRepLevels}, several algorithms have been proposed to find the best codebook. For instance, in the original paper \cite{marcellin_TCQ} an iterative algorithm is given for alphabet optimization that improves on the uniform distribution. In other approaches, in an attempt to achieve some of the \emph{boundary} gain too \cite{eyuboglu_LatticeTrellisQuant}, the occurrences of each reconstruction level are taken into account during the optimization (and the coding) procedure. Examples are given in \cite{yang_trellisMemoryless} (that refines the work in \cite{laroia_tb-svq}) and in \cite{marcellin_OnECTCQ}, where the concept is more elegantly approached via Lagrangian optimization.

Observing that the size of the reproduction alphabet puts (at rates close to $\log_2|\mathcal{A}|$, with $|\mathcal{A}|$ the size of the alphabet) a non-negligible limit on the achievable rate-distortion performance \cite{finamore_EncodingFiniteOutput}, recent approaches \cite{vanDerVleuten_TCQconst,eriksson_lc-tsc-tcom} have investigated the opportunity to use larger codebooks for improved performance. In particular, in \cite{eriksson_lc-tsc-tcom} a linear congruential recursion is used to randomize with a very small computational burden the labels associated with the trellis branches.

The performance improvement over the basic TCQ systems was eventually obtained by taking into account for the rate (i.e.~by entropy coding) or by increasing the reproduction set, but essentially keeping the trellis structure of a convolutional binary code for codeword randomization. As observed in \cite{eriksson_lc-tsc-tcom}, in fact, there seems to be no performance gain from changing the very simple shift-register transitions.

Despite the fact that eventually TCQ heavily relies on convolutional codes, the current know-how about the algebraic properties of the convolutional codes has not been considered for the design of TCQ systems, which are instead based on ad-hoc convolutional codes (such as the ones in \cite{ungerboeck_ChCodWithMultilevelSignals}). In this letter a TCQ design technique based on maximum-Hamming-distance convolutional codes is investigated for high rate TCQ. The obtained codes are shown to be competitive with the ones in the literature for the same computational complexity.\footnote{This fact seems to show that the folk channel coding result stating essentially that all ``sensible'' labelings lead to nearly equivalent performance \cite{wesel_constLabeling} holds as well in the source coding domain.}

\section{Maximum-Hamming-Distance-Based {\TCQr} Design}\label{s:design}
Suppose that the source to be quantized produces sequences of independent and identically distributed samples. According to the generally accepted Gersho's conjecture \cite{gersho_AsymptOptBlockQuant}, the Voronoi cells of the best vector quantizer are all asymptotically congruent to the same polytope, with the possible exception of the ones touching the boundary of the typical set over which the source outcomes are uniformly distributed. At high rates, i.e.~when most Voronoi cells lie in the interior of the typical set, it is then reasonable that all \emph{granular} gain \cite{eyuboglu_LatticeTrellisQuant} can be achieved by means of a \emph{geometrically uniform code} \cite{forney_GeomUnifCodes}. Successively, entropy coding achieves all \emph{boundary} gain \cite{eyuboglu_LatticeTrellisQuant}, independently of the source distribution.

If the extended codebook $S\subset\mathbb{R}^N$ is a \emph{geometrically uniform set} that can be partitioned into the cosets of $S'\lhd S$, and $S/S'$ is isomorphic to the group $(\{0,1\}^2,\oplus)$, then TCQ defines a geometrically uniform code \cite{forney_GeomUnifCodes}. In particular, it is the \emph{generalized coset code} $\mathcal{C}(S/S',\mathbf{C}_{L-1})\subset\mathbb{R}^{NL}$, where $\mathbf{C}_{L-1}$ are the codewords of maximum \emph{degree} $L-1$ of the used $(2,1)$ convolutional code \cite{mceliece_ConvCodes_inbook}. The rate-distortion performance of $\mathcal{C}(S/S',\mathbf{C}_{L-1})$ depends on the reciprocal position of the \emph{coset representatives} of the partition $\mathcal{C}(S/S',\mathbf{C}_{L-1})/{S'}^L$ which are the closest to the null element $0_{S'}^L$ of ${S'}^L$. The aim is hence to \emph{maximize the reciprocal distance of these coset representatives}.

\subsection{Distance Preserving Labelings}
Since the representatives of the different cosets are related to the different paths on the trellis, and then to different codewords of the underlying convolutional code, it is worth to investigate if there exist systems that associate \emph{more} distant coset representatives (in Euclidean signal space) to \emph{more distant} codewords (i.e.~with large {\hd}). In such a case, distance-optimal binary convolutional codes would lead to generalized coset codes with good distance properties.

Any coset representative of the partition $\mathcal{C}(S/S',\mathbf{C}_{L-1})/{S'}^L$ is simply an $L$-tuple $(s_0,s_1,\dots,s_{L-1})$ of points of $S$ such that $s_n\in \mu(a_n)$, $n=0,1,\dots,L-1$, where $\mu$ is the \emph{isometric labeling} \cite{forney_GeomUnifCodes} of the partition $S/S'$ and $(a_0,a_1,\dots,a_{L-1})$ is the codeword of the underlying convolutional code which specifies the considered coset. Since the (squared) {\ed} is additive, among the possible coset representatives of a coset the closest to the point $0_{S'}^L$ is the one for which $s_n\in \mu(a_n)$ is the closest to $0_{S'}$, for each $n=0,1,\dots,L-1$. Let us denote with $\bar{S}$ the subset of $S$ which exactly contains all the coset representatives of the partition $S/S'$ which are the closest to $0_{S'}$.

Like (squared) {\ed}, {\hd} is additive. To assign more distant coset representatives to more distant codewords it is then sufficient for $\mu$ to label the elements of $\bar{S}$ (and hence of $S$) with binary tuples such that the closer two points are in $N$-dimensional Euclidean space, the smaller is the {\hd} between the corresponding labels. A binary isometric labeling satisfying this property has been called a \emph{distance preserving labeling}. In the following common cases such a labeling exists.

\begin{figure}%
\centering
\subfigure[]{
\includegraphics[scale=1]{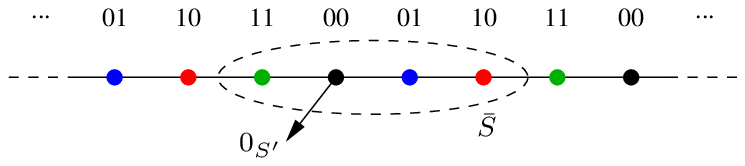}
\label{f:isolabel1}
}\\
\subfigure[]{
\includegraphics[scale=1]{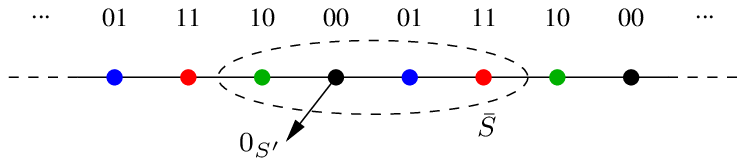}
\label{f:isolabel2}
}\\
\subfigure[]{
\includegraphics[scale=1]{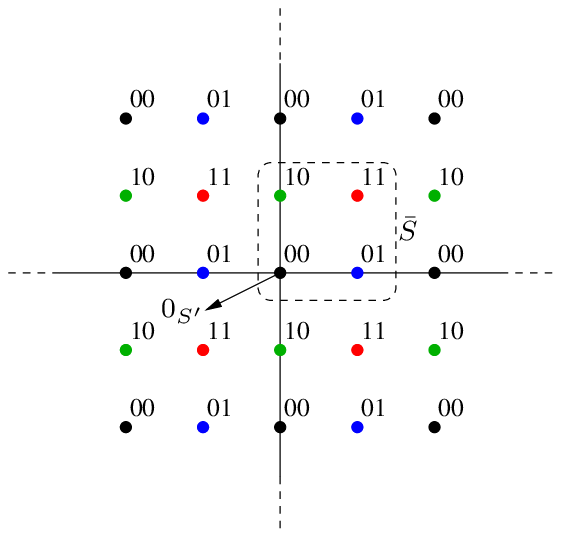}
\label{f:isolabel2_2d}
}
\caption{(a,b) Two isometric labelings for the partition $\mathbb{Z}/4\mathbb{Z}$; (c) a \emph{distance preserving labeling} for the partition $\mathbb{Z}^2/2\mathbb{Z}^2$.}
\end{figure}

\begin{itemize}
  \item One-Dimensional Partition $\mathbb{Z}/4\mathbb{Z}$. Consider, for $N=1$, the partition $S/S'=\mathbb{Z}/4\mathbb{Z}$, that corresponds to the (scaled and) infinitely extended alphabet case obtained from the scalar example in \cite{marcellin_TCQ}. Usually, the partition is labeled as in Fig.~\ref{f:isolabel1} and ad-hoc $(1,2)$ binary convolutional codes are used. A possible distance preserving labeling is shown in Fig.~\ref{f:isolabel2}: labels differing by one or two bits are respectively given to cosets whose minimum (squared) {\ed} equals one or four.
  \item Two-Dimensional Partition $\mathbb{Z}^2/2\mathbb{Z}^2$. For $N=2$, consider the partition $S/S'=\mathbb{Z}^2/2\mathbb{Z}^2$, that again corresponds to the (scaled and) infinitely extended alphabet case obtained from the {\td} example in \cite{marcellin_TCQ}. A distance preserving labeling is shown in Fig.~\ref{f:isolabel2_2d}: labels differing by one or two bits are respectively given to cosets whose minimum (squared) {\ed} equals one or two.
\end{itemize}

\section{Simulation Results}\label{s:results}
At high rates, and assuming that the symbols output by the {\tcqr} are entropy-coded down to their entropy, the asymptotic performance of \emph{geometrically uniform} TCQ is measured by the asymptotic \emph{normalized second moment} (per dimension) of the Voronoi cell of $\mathcal{C}(S/S',\mathbf{C}_{L-1})$, independently from both the rate and the probability distribution of the source \cite{eyuboglu_LatticeTrellisQuant}. Equivalently, the performance can be expressed in terms of coding gain w.r.t.~entropy-coded uniform quantization by the asymptotic value of the \emph{granular gain}, $\gamma_g$, which simply represents the factor of reduction of the normalized second moment obtained by \emph{shaping} an hypercube into the shape of the Voronoi cell of the {\tcqr}.

In the following, hence, the granular gain of the traditional TCQ systems found in the literature with the granular gain of maximum-{\hd}-based systems having the same number of states (i.e., exactly the same computational complexity) are compared.

\subsection{Experimental Setup}
The \emph{approximated} second moment of a Voronoi cell can be evaluated only by simulation. If an $LN$-dimensional region containing several cells (along each coordinate) is uniformly populated by $N_v$ random vectors (sequences), the average per dimension energy of the corresponding \emph{quantization errors} approximates the second moment (per dimension) $P(\mathcal{C})$ itself.\footnote{In fact, the great number of cells guarantees that there are no border effects in this evaluation.} Hence, the quantity
\begin{equation}
\label{e:tildeP}
\tilde{P}(\mathcal{C}) = \frac{1}{N_v}\sum_{i=0}^{N_v-1}\left[\frac{1}{LN} \sum_{j=0}^{LN-1} (x_{i,j}-\tilde{x}_{i,j})^2 \right]
\end{equation}
can be evaluated, where $x_{i,j}$ is the $j$-th realization of the $i$-th random sequence (and $\tilde{x}_{i,j}$ is its reconstruction). The fidelity of this approximation can be measured as follows. Note that for high values of $LN$, one can appeal to the \emph{central limit theorem} 
and hence assume that the expression in square brackets in (\ref{e:tildeP}) is a Gaussian r.v.
~belonging to $\mathcal{N}(P(\mathcal{C}),\sigma^2)$. Consequently, $\tilde{P}(\mathcal{C})\in \mathcal{N}(P(\mathcal{C}),\sigma^2/N_v)$ and in the $95$\% of the cases
\begin{equation}
\left| \tilde{P}(\mathcal{C}) - P(\mathcal{C})\right| \leq 2\frac{\sigma}{\sqrt{N_v}} \cong 2\frac{\bar{\sigma}}{\sqrt{N_v}} \triangleq \Delta P\;,\nonumber
\end{equation}
where the best \emph{unbiased} estimate of the standard deviation given the $N_v$ measured distortion samples can be used as $\bar{\sigma}$.

In the experiments, the granular gain is evaluated with $N_v=5000$ sequences of length $LN=1000$, randomly distributed in the hypercube with sides equal to $2^R\cdot 2$ and to $2^R\cdot\sqrt{2}$ respectively for the case $\mathbb{Z}/4\mathbb{Z}$ and the case $\mathbb{Z}^2/2\mathbb{Z}^2$, with $R=8$.

\subsection{Results at High Rates}
The top-half of Table \ref{t:1d} compares the granular gains of the $\mathbb{Z}/4\mathbb{Z}$-TCQ systems in \cite{marcellin_TCQ} with the ones of the proposed systems that are based on the labeling of Fig.~\ref{f:isolabel2}. For each number of states, the codes are shown in octal form. The codes shown in the fourth column of Table \ref{t:1d} are distance-optimal $(2,1)$ convolutional codes taken from \cite{mceliece_ConvCodes_inbook}. The values in the left hand side of Table \ref{t:1d} confirm the results at high rates of \cite{marcellin_TCQ} and \cite{eyuboglu_LatticeTrellisQuant}. 

\begin{table}%
\renewcommand{\arraystretch}{1.3}
\caption{Asymptotic granular gain (in dB) for the partitions $\mathbb{Z}/4\mathbb{Z}$ (top-half) and $\mathbb{Z}^2/2\mathbb{Z}^2$ (bottom-half); $LN=1000$ samples.}
\label{t:1d}
\centering
\begin{tabular}{c|c|c|c|c}
\hline
states & \multicolumn{2}{c|}{Ungerboeck codes} & \multicolumn{2}{c}{distance-optimal codes} \\
       & code & granular gain                  & code & granular gain\\
\hline\hline
$4$ & $[5\ 2]$ & $1.005\pm .0018$       & $[5\ 7]$ & $1.003\pm .0018$\\ 
\hline
$8$ & $[13\ 04]$ & $1.097\pm .0016$     & $[13\ 17]$ & $1.098\pm .0016$\\ 
\hline
$16$ & $[23\ 04]$ & $1.162\pm .0014$    & $[23\ 35]$ & $1.169\pm .0014$\\ 
\hline
$32$ & $[45\ 10]$ & $1.232\pm .0013$    & $[53\ 75]$ & $1.233\pm .0013$\\ 
\hline
$64$ & $[103\ 024]$ & $1.282\pm .0012$  & $[133\ 171]$ & $1.290\pm .0012$\\ 
\hline
$128$ & $[235\ 126]$ & $1.332\pm .0010$ & -- & --\\ 
\hline                                               
$256$ & $[515\ 362]$ & $1.369\pm .0010$ & $[561\ 753]$ & $1.375\pm .0010$\\ 
\hline                                               
$1024$ & -- & --                        & $[2335\ 3661]$ & $1.428\pm .0008$\\ 
\hline\hline
$4$ & $[5\ 2]$ & $0.981\pm .0018$       & $[5\ 7]$ & $0.980\pm .0019$\\
\hline
$8$ & $[13\ 04]$ & $1.074\pm .0017$     & $[13\ 17]$ & $1.074\pm .0017$\\
\hline
$16$ & $[23\ 04]$ & $1.124\pm .0016$    & $[23\ 35]$ & $1.137\pm .0015$\\
\hline
$32$ & $[45\ 10]$ & $1.171\pm .0015$    & $[53\ 75]$ & $1.191\pm .0015$\\
\hline
$64$ & $[103\ 024]$ & $1.225\pm .0014$  & $[133\ 171]$ & $1.243\pm .0014$\\
\hline
$128$ & $[235\ 126]$ & $1.272\pm .0014$ & -- & --\\
\hline
$256$ & $[515\ 362]$ & $1.303\pm .0013$ & $[561\ 753]$ & $1.311\pm .0013$\\
\hline
$1024$ & -- & --                        & $[2335\ 3661]$ & $1.351\pm .0013$\\
\hline
\end{tabular}
\end{table}

For most number of states, the proposed codes exactly achieve the same granular gain. In particular, the codes with $4$ and $8$ states lead to systems exactly equivalent to the corresponding ones based on Ungerboeck codes. A measurable, yet almost negligible improvement is instead measured for codes with $16$, $64$, and $256$ states.

The $1024$-states code $[2335\ 3661]$ is asymptotically only about $0.1$ dB farther from the \emph{ultimate granular gain} of $1.53$ dB that separates \emph{entropy constrained scalar quantization} from the \emph{Shannon lower bound}. This means that this {\tcqr}, at high rates, almost fills the gap between the performance of scalar quantization and the theoretic rate-distortion curve.

The $\mathbb{Z}^2/2\mathbb{Z}^2$-based TCQ systems are compared in the bottom-half of Table \ref{t:1d}. In these cases each successive couple of samples is constrained on a squared lattice (a \emph{lattice translate} of $2\mathbb{Z}^2$), and the convolutional codes can only randomize the relative position of the successive couples of samples. Consequently, compared to the $\mathbb{Z}/4\mathbb{Z}$ case, where the convolutional code offers the possibility to control the relative position of each successive sample, giving more freedom to the system, the overall performance is slightly reduced.

\begin{figure}%
\centering
\includegraphics[scale=.65]{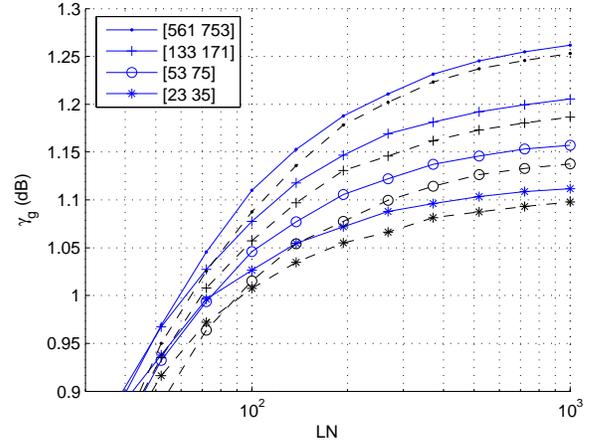}
\caption{Experimental granular gain for the partition $\mathbb{Z}^2/2\mathbb{Z}^2$, using $16$-, $32$-, $64$- and $256$-state distance-optimal codes. The dashed lines represent the granular gain obtained using the Ungerboeck codes with the same number of states.}
\label{f:typ1MDL_mdb}
\end{figure}

However, improvements are again obtained over the systems usually found in literature. The asymptotic coding gain is in fact higher for all the codes with $16$, $32$, $64$, and $256$ states. Close examination of the granular gain of these codes for different numbers of coded samples, compared with the granular gain of the traditional systems using the corresponding codes (see Fig.~\ref{f:typ1MDL_mdb}), shows that they give improvements even for $LN<1000$.

\subsection{Results at Low Rates}
Since the asymptotic gains describe only the high rate performance, it is interesting to investigate if the proposed codes are competitive with the traditional ones even at low rates. By simply limiting the extended codebook to only $2^{R+1}$ consecutive elements of $0.5+\mathbb{Z}$ (assuming a symmetric source distribution with respect to zero) it would be possible to achieve any (integer) desired limited bit rate. However, the performance increases if such a finite alphabet is optimized according to some training data set, while the encoding complexity remains essentially the same.

The comparison, relative to the optimized finite alphabets with $R$ equal to $1$, $2$, and $3$ bit/sample, is shown in Table \ref{t:finite_rate} 
for both the uniform and the Gaussian distribution.

\begin{table}%
\renewcommand{\arraystretch}{1.3}
\centering
\caption{Comparison of the signal-to-quantization noise ratio relative to the traditional and the proposed codes (using $M$ states), for $\mathbb{Z}/4\mathbb{Z}$-TCQ; $LN=1000$ uniformly (top-half) and Gaussian (bottom-half) distributed samples.}
\label{t:finite_rate}
\begin{tabular}{c|c|c|c}
\hline
rate    & $M$ & \multicolumn{2}{c}{SQNR (dB)} \\
bit/sample & & traditional codes & proposed codes \\
\hline
\hline
  & 16  & $6.405\pm .0029$ & $6.412\pm .0029$\\
1 & 64  & $6.496\pm .0028$ & $6.519\pm .0027$\\ 
  & 256 & $6.579\pm .0027$ & $6.593\pm .0026$\\
\hline
  & 16  & $12.802\pm .0023$ & $12.812\pm .0023$\\
2 & 64  & $12.915\pm .0021$ & $12.936\pm .0021$\\ 
  & 256 & $13.012\pm .0019$ & $13.022\pm .0019$\\
\hline
  & 16  & $19.024\pm .0019$ & $19.033\pm .0019$\\
3 & 64  & $19.142\pm .0017$ & $19.157\pm .0017$\\ 
  & 256 & $19.236\pm .0015$ & $19.243\pm .0015$\\
\hline
\hline
  & 16  & $5.285\pm .0060$ & $5.310\pm .0061$\\
1 & 64  & $5.440\pm .0060$ & $5.478\pm .0060$\\ 
  & 256 & $5.570\pm .0059$ & $5.583\pm .0059$\\
\hline
  & 16  & $10.778\pm .0071$ & $10.796\pm .0072$\\
2 & 64  & $10.927\pm .0072$ & $10.948\pm .0072$\\ 
  & 256 & $11.035\pm .0072$ & $11.043\pm .0072$\\
\hline
  & 16  & $16.412\pm .0087$ & $16.431\pm .0085$\\
3 & 64  & $16.560\pm .0082$ & $16.578\pm .0081$\\ 
  & 256 & $16.673\pm .0087$ & $16.683\pm .0087$\\
\hline
\end{tabular}
\end{table}

The results show a similar improvement with respect to the TCQ systems in \cite{marcellin_TCQ} for the codes with $16$, $64$, and $256$ states. Distance-optimal convolutional codes are hence able to distribute the reconstruction values in a maximal-distance fashion even in non-regular arrays. If the burden of randomizing the labels of the trellis and increasing the codebook size is not wanted (see \cite{vanDerVleuten_TCQconst,eriksson_lc-tsc-tcom}), then the proposed systems are a feasible alternative to traditional {\tcqr}s.

\section{Conclusion}\label{s:conclusion}
In this letter an approach to TCQ design has been presented that is based on coupling a \emph{distance preserving labeling} of the extended codebook with maximum-{\hd} convolutional codes. In this way, the relation between the minimum distance of the feasible reconstructed sequences and the {\hd} of the underlying convolutional code is exposed. 

Results show that it is possible and somewhat advantageous to conduct the search for good trellis codes into the binary domain, i.e.~to look for \emph{good} convolutional codes. In fact, the codes found are competitive with or somewhat better than the ones in literature. In the $\mathbb{Z}/4\mathbb{Z}$ case better codes have been found for the $16$, $64$, and $256$ states case, while in the $\mathbb{Z}^2/2\mathbb{Z}^2$ case better codes have been found for the $16$, $32$, $64$, and $256$ states cases. In addition, a $1024$ states code is given that, according to the authors' knowledge, was never published before in the context of TCQ.

\bibliographystyle{IEEEtran.bst}
\bibliography{IEEEabrv,nonIEEEabrv,refs_TCQ,refs_DSC,refs_books}

\end{document}